\documentclass[showpacs,twocolumn,preprintnumbers,amsmath,amssymb,prb,superscriptaddress]{revtex4}

\usepackage[sort&compress]{natbib}
\usepackage{graphicx}
\usepackage{dcolumn}
\usepackage{bm}

\newcommand{\cobaltite}{Co$_3$O$_4$}
\newcommand{\magnetite}{Fe$_3$O$_4$}
\newcommand{\spinel}{MgAl$_2$O$_4$}
\newcommand{\sapphire}{$\alpha$-Al$_2$O$_3$}
\newcommand{\oC}{$^\mathrm{o}$C}

\begin{document}


\title{Experimental study of the interfacial cobalt oxide in \cobaltite/\sapphire(0001) epitaxial films}

\author{C. A. F. Vaz}
\email[Corresponding author. Email: ]{carlos.vaz@yale.edu}%
\affiliation{Department of Applied Physics, Yale University, New
Haven, Connecticut 06520}%
\affiliation{Center for Research on Interface Structures and
Phenomena (CRISP), Yale University, New Haven, Connecticut 06520}%

\author{D. Prabhakaran}
\affiliation{Department of Physics, Clarendon Laboratory, Oxford University, Oxford OX1 3PU, United Kingdom}%

\author{E.~I. Altman}
\affiliation{Department of Chemical Engineering, Yale University,
New Haven, Connecticut 06520}%
\affiliation{Center for Research on Interface Structures and
Phenomena (CRISP), Yale University, New Haven, Connecticut 06520}%

\author{V.~E. Henrich}
\affiliation{Department of Applied Physics, Yale University, New
Haven, Connecticut 06520}%
\affiliation{Center for Research on Interface Structures and
Phenomena (CRISP), Yale University, New Haven, Connecticut 06520}%

\date{\today}

\begin{abstract}
A detailed spectroscopic and structural characterization of
ultrathin cobalt oxide films grown by O-assisted molecular beam
epitaxy on \sapphire(0001) single crystals is reported. The
experimental results show that the cobalt oxide films become
progressively more disordered with increasing thickness, starting
from the early stages of deposition. Low energy electron diffraction
patterns suggest that the unit cell remains similar to that of
\sapphire(0001) up to a thickness of 17 \AA, while at larger
thicknesses a pattern identified with that of \cobaltite(111)
becomes visible. X-ray photoelectron spectroscopy reveals sudden
changes in the shape of the Co 2p lines from 3.4 to 17 \AA\ cobalt
oxide thickness, indicating the transition from an interfacial
cobalt oxide layer towards [111]-oriented \cobaltite. In particular,
the absence of characteristic satellite peaks in the Co 2p lines
indicates the formation of a trivalent, octahedrally coordinated,
interfacial cobalt oxide layer during the early stages of growth,
identified as the Co$_2$O$_3$ corundum phase.
\end{abstract}



\maketitle

\section{Introduction}

The stability of polar surfaces, characterized by a net surface
charge, has long been a topic of much interest, since the large
electrostatic energies associated with such surfaces are expected to
lead to modified electronic and atomic structures, with attendant
changes in physical properties.\cite{HC94,Noguera00,GFN08} This
often leads to surface structures that depart significantly from a
simple truncation of the bulk crystal, exhibiting reconstructions,
faceting, surface roughening, altered
valencies,\cite{HKWG78,Tasker79,RW99,Noguera00,JPS+02,LPP+05,GFN08,MBG+08,VWA+09,VHAA09}
and that can give rise to other exotic phenomena, such as the onset
of two-dimensional metallic states in LaAlO$_3$/SrTiO$_3$(001)
interfaces.\cite{SS08,CKB09}

Recently, we reported the epitaxial growth of
\cobaltite(110)/\spinel(110) and \cobaltite(111)/\sapphire(0001)
thin films, which were found to exhibit ($1 \times 1$) surfaces,
despite the fact that both surfaces are polar; while the as-grown
film surfaces show some degree of disorder, annealing in air results
in atomically smooth films for \cobaltite(110), and improved
morphology for the \cobaltite(111) films, while retaining the ($1
\times 1$) surface structure.\cite{VWA+09,VHAA09} The stability of
these surfaces was attributed to a modified surface valency of the
Co cations, corresponding effectively to a surface inversion in the
spinel structure; identical conclusions were reached in a study of
the growth of twinned ($1\times 1$) [111]-oriented \cobaltite\ films
on Ir(001)-($1\times 1$),\cite{MBG+08} and in
\magnetite(111)/Pt(111) thin films, which also exhibit a ($1 \times
1$) surface.\cite{RW99} One unresolved issue remains the interface
structure at the early stages of growth of
\cobaltite(111)/\sapphire(0001). The observation of a significant
amount of disorder occurring at the \cobaltite/\sapphire(0001)
interface was tentatively attributed to the possibility of the
formation of an off-stoichiometric cobalt oxide, perhaps closer to
the corundum Co$_2$O$_3$ phase. In this report, we present the
results of a detailed study of the early stages of growth of
\cobaltite/\sapphire(0001) thin films, aiming at understanding the
electronic and crystal surface structure of the interfacial oxide
layer. We show, from x-ray photoelectron spectroscopy of the Co 2p
edge, that the cobalt oxide growth begins with the formation of a
Co$^{3+}$-rich interfacial layer, which we associate with the cobalt
sesquioxide Co$_2$O$_3$, crystalizing in the corundum
structure.\cite{CJM71,Samsonov82}

This is a surprising result in that the cobalt sesquioxide,
Co$_2$O$_3$, is not the preferred cobalt oxide at the temperatures
and pressures attainable in molecular beam epitaxy growth
conditions. In fact, the electronic and catalytic properties of
Co$_2$O$_3$ have not been studied extensively, possibly due to
difficulties in synthesizing this compound. Chenavas et
al.\cite{CJM71} suggest the existence of a high pressure phase with
low spin Co$^{3+}$ ($a=4.882$ \AA, $c=13.38$ \AA) and a low pressure
phase, with high spin Co$^{3+}$ ($a=4.782$ \AA, $c=12.96$ \AA) based
on the observation of a reduction in the unit cell volume upon
annealing in air at 400\oC\ for 30 min, where the spin state was
inferred from the smaller ionic radius of high spin Co$^{3+}$. {\it
Ab initio} (ground state) calculations for corundum Co$_2$O$_3$
suggest also that this oxide is energetically stable.\cite{CS97} The
most stable oxides of cobalt include the high temperature rocksalt
CoO (cobaltous oxide) phase, where the Co$^{2+}$ ($S=3/2$) occupy
octahedral sites, and the mixed valence cobalt cobaltite,
Co$_3$O$_4$, crystalizing in the normal spinel structure ($a=8.086$
\AA)\cite{PBBC80} with the Co$^{2+}$ occupying tetrahedral sites and
the Co$^{3+}$ occupying octahedral
sites.\cite{Cossee58,Roth64,SH73,IAK+82} The octahedral Co$^{3+}$
ions are in a low spin state, $S=0$, while the tetrahedral Co$^{2+}$
ions are in a high spin state ($S=3/2$), which order
antiferromagnetically below about 40 K.\cite{Roth64,SH76}

\section{Experimental}

Film growth was carried out in an ultrahigh vacuum (UHV) oxide
molecular beam epitaxy deposition system (base pressure of $1\times
10^{-9}$ mbar), using conditions identical to those reported
earlier.\cite{VHAA09,VWA+09} Prior to film growth, the \sapphire\
substrate was annealed at 600\oC\ in UHV for 60 min, followed by
exposure to atomic oxygen at 300\oC\ for 30 min. This procedure
results in \sapphire(0001) surfaces free of carbon and yielding
sharp reflection high energy electron diffraction (RHEED) and low
energy electron diffraction (LEED) patterns, characteristic of
atomically smooth surfaces (see Fig.~\ref{fig:RHEED} and
Fig.~\ref{fig:LEED}). The only impurities detected by Auger electron
spectroscopy (AES) consist of trace amounts of Ca (2\%) and K
(0.2\%). Film growth was carried out at 300\oC\ by simultaneous
exposure of the substrate to an atomic Co beam evaporated thermally
from an effusion cell and to an atomic oxygen beam generated by a
magnetron plasma source; the O$_2$ partial pressure during growth
was set to $3\times 10^{-5}$ mbar. The Co deposition rate was about
1 \AA/min, as measured by a calibrated quartz thickness monitor.
Growth rates were monitored before and after deposition throughout
this study. The film growth was interrupted at several stages of the
deposition process for LEED, AES and x-ray photoelectron
spectroscopy (XPS) analysis, namely, after deposition of 1, 2, 10,
20 and 70 \AA\ Co. From {\it ex situ} x-ray reflectometry of the
film carried out after growth, the oxide film thickness was found to
be $120\pm 10$ \AA, where the error bar includes possible systematic
errors in the measurement. From this value, the corresponding cobalt
oxide film thicknesses are obtained: 1.7, 3.4, 17, 34, and 120 \AA.
LEED, AES and XPS measurements were performed after transferring the
sample under UHV from the growth chamber to a dedicated analysis
chamber with a base pressure of $1\times 10^{-10}$ mbar; typical XPS
measurement times ranged from 2-5 h. Before continuation of the film
growth, the film surface was exposed to the atomic oxygen beam for 5
min with the sample held at 300\oC. {\it Ex situ} x-ray diffraction
(XRD) measurements were performed on the 120 \AA\ film on a Shimadzu
diffractometer operating in the parallel beam optics geometry.

\section{Results and discussion}

\begin{figure}[t!bh]
\begin{centering}
\includegraphics*[width=6.8cm]{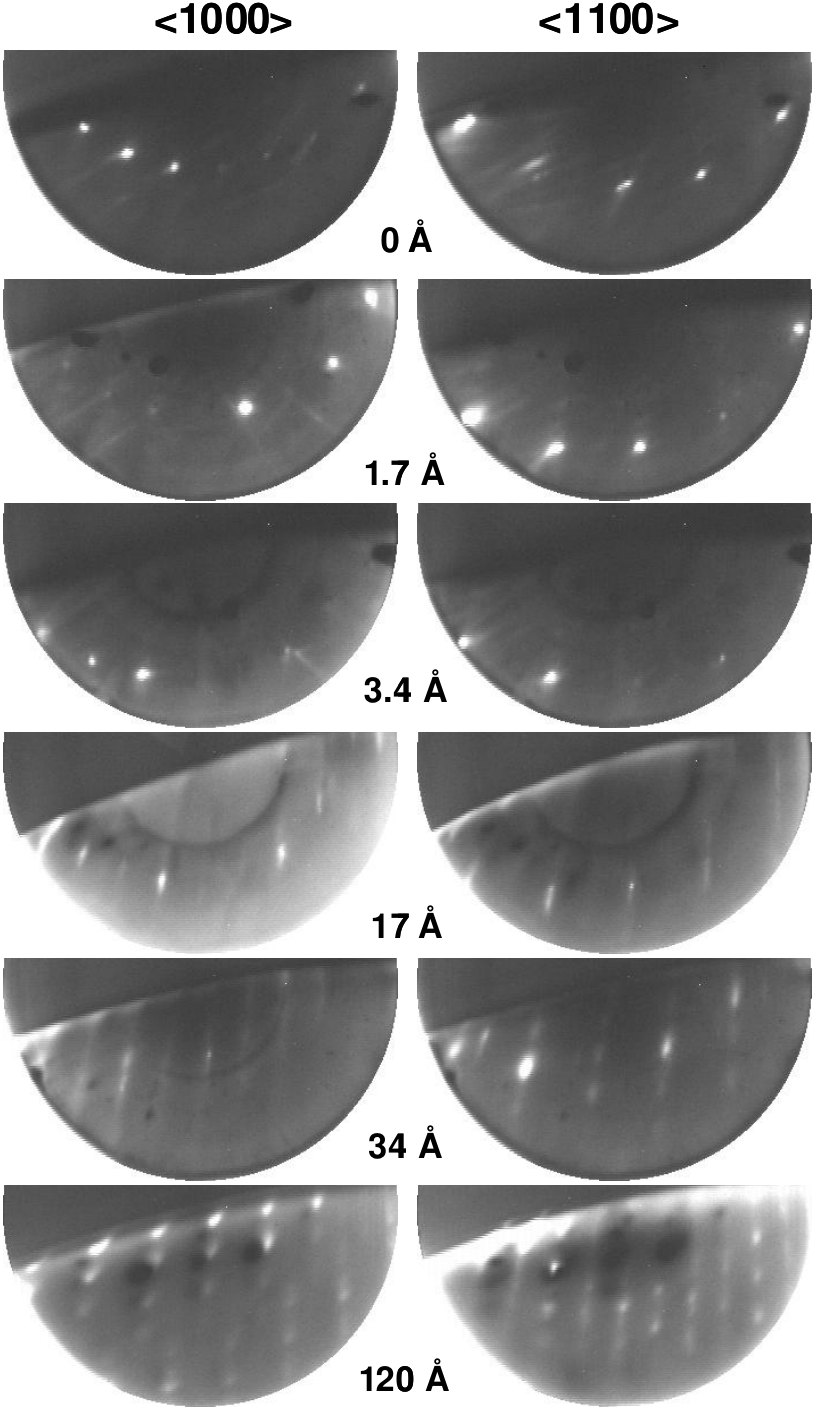}
\caption{RHEED patterns of the cobalt oxide film along two different
azimuths of the \sapphire(0001) surface at several stages of the
film growth, as labeled. The incident electron beam energy is 15
keV.} \label{fig:RHEED}
\end{centering}
\end{figure}

Film crystallinity was monitored during growth using RHEED, and the
diffraction patterns after completion of each layer are shown in
Fig.~\ref{fig:RHEED}. As a general trend, the diffuse background
scattering increased with increasing deposition thickness; the
diffraction spots broadened with film thickness, starting with sharp
patterns at low coverages (1.7, 3.4 \AA), which became streakier at
intermediate thicknesses (17, 34 \AA) and spotty, transmission-like,
at larger thicknesses, in agreement with previous
observations.\cite{VHAA09} These results suggest that the cobalt
oxide growth proceeds in the Stranski-Krastanov mode.\cite{VSH84}

\begin{figure}[t!]
\begin{centering}
\includegraphics*[width=8.5cm]{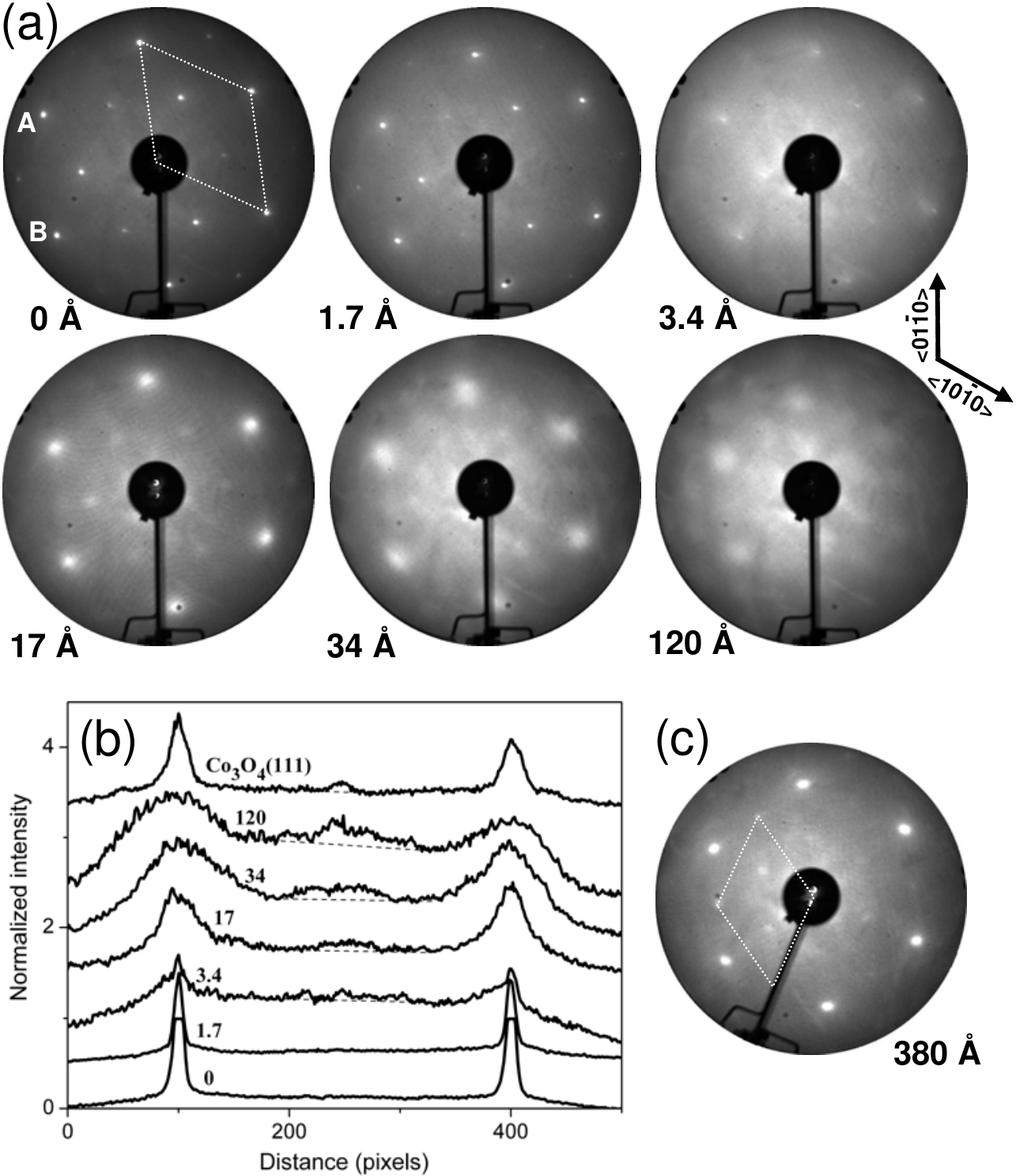}
\caption{(a) LEED patterns of the cobalt oxide film at several
stages of film growth, as labeled. The incident electron beam energy
is 100.0 eV for the patterns shown, except for the 34 \AA\ (116.8
eV) and 120 \AA\ films (138.2 eV). (b) LEED line profiles across the
diffraction spots labeled A and B in (a) for the different cobalt
oxide thickness; the profile for \cobaltite(111) corresponds to a
LEED pattern of a 38 nm \cobaltite(111)/\sapphire(0001) film
annealed in air at 600\oC\ for 14 h (104.3 eV), shown in (c). The
dashed rhombus in (a) and (c) represent the LEED unit cells of
\sapphire\ and \cobaltite, respectively.} \label{fig:LEED}
\end{centering}
\end{figure}

LEED patterns taken at room temperature immediately after completion
of each layer are shown in Fig.~\ref{fig:LEED}, showing ($1\times
1$) diffraction patterns whose overall features agree with the RHEED
results. The oblique cell drawn in Fig.~\ref{fig:LEED} (0 \AA)
corresponds to the \sapphire(0001) unit cell (corundum, $a = 4.7570$
\AA, $c = 12.9877$ \AA\ in the hexagonal
representation);\cite{NH62,Cousins81,KE90} the three-fold symmetric
LEED pattern of the \sapphire(0001) surface indicates that it is
composed predominantly of double-layer atomic steps. The LEED
pattern symmetry remains similar to that of the substrate up to 17
\AA, while for 34 \AA\ its starts resembling that of
\cobaltite(111).\cite{PMCL08,VHAA09} This evolution in the LEED
patterns can also be followed in Fig.~\ref{fig:LEED}(b), which shows
the line profiles across the spots labeled A and B in
Fig.~\ref{fig:LEED}. In order to correct for charging, slight
differences in sample positioning, and for the different electron
beam energies, the distance between these spots was normalized to
the same value for all thicknesses. While for \sapphire(0001) no
features are present between these spots, in \cobaltite(111) there
is an intermediate diffraction spot, the presence of which can be
used to identify the onset of this phase, at about 17 \AA. The XRD
measurements on the 120 \AA\ film show the presence of the ($hhh$)
planes of \cobaltite\ at the angle positions corresponding to the
bulk values, indicating that the film is fully relaxed. The rocking
curve around the (222) plane shows a single gaussian peak, with a
width of 0.026$^\mathrm{o}$, which corresponds to a characteristic
lengthscale in real space of about 180 nm.

Spectroscopic characterization by XPS and AES was performed for each
layer thickness, but while XPS could be performed for all cobalt
oxide thicknesses, AES could only be performed up to 17 \AA, due to
sample charging. The XPS spectra were obtained using the Mg
K$_\alpha$ line ($h\nu = 1253.6$ eV) of a double anode x-ray source
and a double pass cylinder mirror analyzer (PHI 15-255G) set at a
pass energy of 25 eV (energy resolution of about 0.8 eV). The XPS
data, acquired in energy steps of 0.05 eV, were smoothed using a
5-point adjacent averaging and corrected for the Mg K$_\alpha$
satellite. Charging is always a concern for insulating samples; one
common method of calibrating the energy scale against charging is to
use the C 1s line from adventitious hydrocarbons,\cite{CO95,ATB00}
but this process cannot be used in these carbon-free samples. X-ray
photoelectron emission charges the sample positively, while charging
in (electron excited) AES may be of either sign, since the number of
ejected electrons may be larger or smaller than the number of
incident electrons (secondary electron emission, which is strongly
energy dependent).\cite{Henrich73,RJBB86,SS00,RFA+08} Hence,
comparing both the XPS and AES data can provide some information
about the extent of charging.

\begin{figure}[t!bh]
\begin{centering}
\includegraphics*[width=8.5cm]{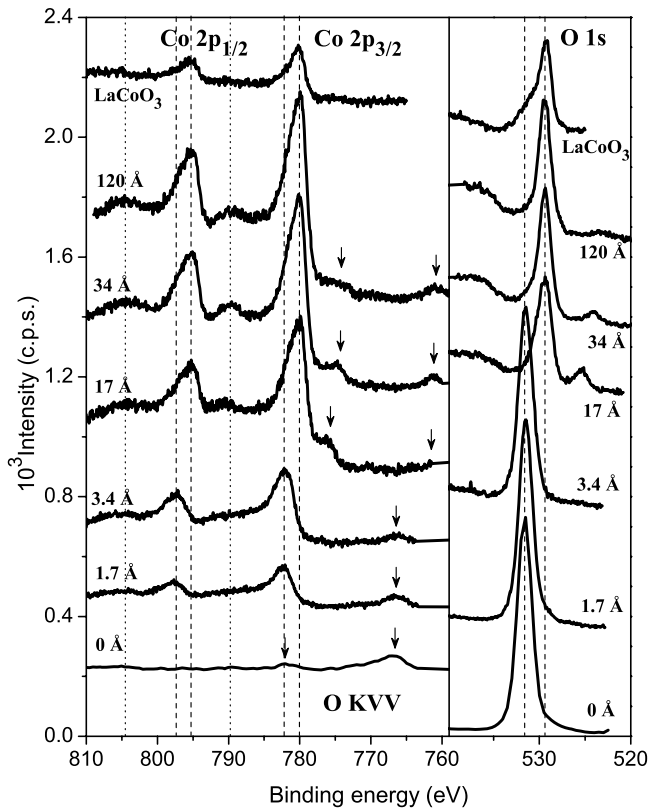}
\caption{XPS spectra as a function of cobalt oxide film thickness at
the Co 2p and O 1s edges. Arrows indicate the energy position of the
O KVV loss peaks, dashed lines indicate main peak positions, and
dotted lines indicated the energy position of the \cobaltite\
satellite peaks. Data have been shifted vertically for convenient
display.} \label{fig:XPS}
\end{centering}
\end{figure}

To correct for sample charging in XPS, we consider the Auger and
photoelectron spectral features of oxygen, namely, the O 1s peak and
the KVV Auger and energy loss peaks, which lie near the Co 2p edge.
We set the O 1s peak of \sapphire\ to the tabulated value of 531.5
eV,\cite{PHI79,CSK05} and the O 1s peak of the 120 \AA\ film,
expected to be representative of \cobaltite,\cite{VHAA09} to the
tabulated value of 529.4 eV.\cite{CBR76,PL04} For the 1.7 and 3.4
\AA\ films, we assume that most of the O contribution to the
photoelectron spectra arises from the substrate, and we tentatively
set the O 1s peak to the same energy position as that of \sapphire.
We find that the KVV Auger loss peaks of O for these films also
align with those for \sapphire\ and that the energy distance from
the first O loss peak to the Co 2p$_{3/2}$ is 15.9 eV for both 1.7
and 3.4 \AA\ cobalt oxide films. For the thicker films, both first
and second O Auger KVV loss peaks are visible, and shifted to much
lower binding energies relative to the Co 2p$_{3/2}$ peak, by 18.9
eV for the first O loss peak and by 5.6 eV for the second O loss
peak for the 34 and 120 \AA\ films; for the 17 \AA\ film, the shift
is slightly smaller, about 14.4 and 4.2 eV, respectively. Therefore,
we can assign the O 1s peak to that of \cobaltite\ for the 34 and 17
\AA\ films, although for the latter thickness the assignment is less
certain. For reference, we also measured the XPS spectra of
LaCoO$_3$, where the Co cations are all trivalent; since LaCoO$_3$
is conducting at room temperature ($\sigma \sim 0.1\mbox{ }
\Omega^{-1}$cm$^{-1}$),\cite{DJD+86,JHKV08} no charging is expected
and no energy corrections have been applied to this XPS spectrum.
The LaCoO$_3$ data are from a single crystal wafer cut along one
pseudo-cubic (110) plane (mechanically polished to optical flatness
and annealed in air at 600\oC\ for 67.5 hours), and were acquired
after cleaning {\it in situ} in oxygen plasma at 300\oC\ for 30 min.
The LaCoO$_3$ single crystal, grown using the floating zone
method,\cite{PBWP05} is twinned, as shown by Laue and x-ray
diffraction, but is otherwise well ordered. The shoulder on the O 1s
peak, which was very prominent in the XPS spectra of the sample
as-inserted to the analysis system, is due to adsorbed hydroxyl
groups that remain on the surface.\cite{HSU76} The XPS spectra for
all samples thus calibrated are shown in Fig.~\ref{fig:XPS}.

\begin{figure}[t!bh]
\begin{centering}
\includegraphics*[width=5.1cm]{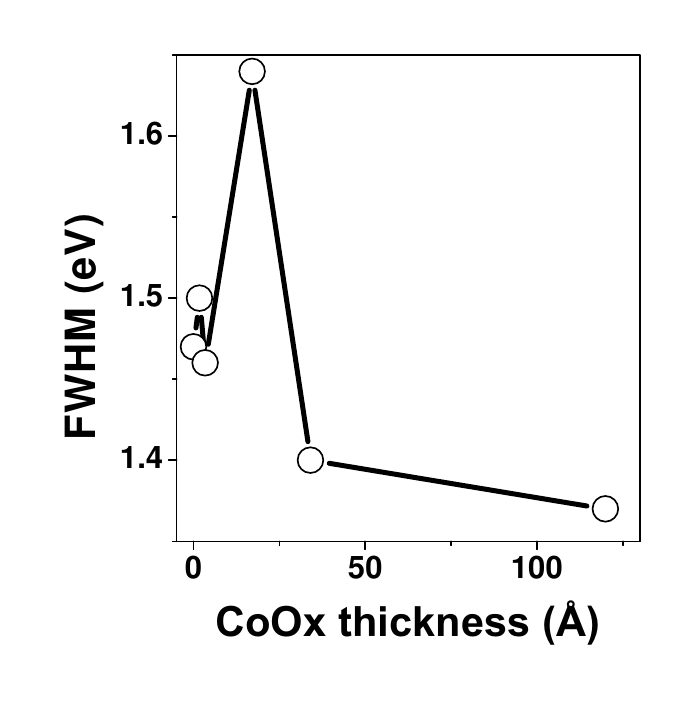}
\caption{Variation of the O 1s XPS peak full width at half maximum
(FWHM) as a function of cobalt oxide thickness.}
\label{fig:XPS_FWHM}
\end{centering}
\end{figure}

Three distinct features in the XPS spectra as a function of cobalt
oxide thickness are apparent: (i) the energy difference between the
2p$_{1/2}$ and 2p$_{3/2}$ peaks remains constant at 15.2 eV with
increasing cobalt oxide thickness, and between \cobaltite,
LaCoO$_3$, and CoO;\cite{PL04} the energy difference between
multiplets is sometimes employed to ascertain the ionic valence
state, but is not suitable for discriminating these different cobalt
oxide compounds. (ii) The higher Co 2p binding energies observed for
the 1.7 and 3.4 \AA\ cobalt oxide films track the change in the
binding energy of the O 1s peak; this energy shift is thus
attributed to band bending effects (pinning of the Fermi level to
that of the \sapphire\ substrate) rather than to a true chemical
shift. This agrees with the fact that for the thinner films (1.7 and
3.4 \AA) we find no splitting or broadening in the O 1s peak, as
shown in Fig.~\ref{fig:XPS_FWHM}; the slightly larger peak width in
the O 1s line at 17 \AA\ could be the result of a small contribution
from the substrate O 1s peak at 531.5 eV. At this thickness, the
substrate contribution becomes very small, as is shown by the fact
that the Al 2p and 2s lines can no longer be detected in the XPS
survey scans (the latter have lower counting statistics than in the
fine Co 2p and O 1s edge energy scans, where a small contribution
may be present). (iii) Significantly, we note the absence of
satellite peaks in the Co 2p spectra of the 1.7 and 3.4 \AA\ films,
and the appearance of weak satellite peaks, characteristic of
\cobaltite,\cite{CBR76,HU77,LAC+99,WHMS04} at larger thicknesses. At
17 \AA, the satellite peaks are still strongly suppressed and lie at
slightly higher binding energies compared to the thicker cobalt
oxide films. This shows that the 1.7 and 3.4 \AA\ films behave
spectroscopically (and electronically) very differently from the
thicker (\cobaltite) films, pointing to a transition from an
interfacial cobalt oxide layer to bulk-like \cobaltite\ at around 17
\AA.

The presence and energy position of satellite peaks in the core and
valence levels of the photoelectron spectra of the 3d transition
metal oxides depend strongly on the ionic environment, cation
valency, and electron
occupancy.\cite{FMW74,KW75,VSC76,BS83,VP85,MUN96} These satellite
peaks arise as a consequence of the fact that several channels are
available that compete for the final (excited) state. In 2p core
level photoelectron emission, the presence of the core-hole leads to
strong modifications in the energy landscape of the cation and anion
(ligand) orbitals that favor screening of the core hole via charge
transfer from $sp$ or ligand orbitals. Two processes compete for the
final state, one corresponding to the case where charge is
transferred from the ligand to the 3d orbital, effectively screening
the core hole, which is usually associated with the main line ({\em
well screened} state, represented by
$|2p^53d^{n+1}\underline{L}\rangle$, where $\underline{L}$
represents the ligand
hole).\cite{Larsson76,LB77,LWHS81,ZWS86,GTE+88,Fujimori88} The other
competing process corresponds to a less well screened state (with
higher apparent binding energy), where no charge transfer from the
ligand takes place and where charge compensation is provided by $sp$
orbitals ({\em unscreened} state, represented by
$|2p^53d^{n}\rangle$). In a simplified cluster
model,\cite{Larsson76,LB77,LWHS81,ZWS86,GTE+88} the photoemission
process is treated as a scattering event from an initial hybridized
state
\begin{equation}
|\psi_g\rangle = \alpha_0|2p^6 3d^{n}\rangle + \alpha_1 |2p^6
3d^{n+1}\underline{L}\rangle + ...
\end{equation}
to final hybridized states of the form:
\begin{equation}
|\psi_f \rangle = \beta_0 |2p^5 3d^{n}\rangle + \beta_1
|2p^53d^{n+1}\underline{L}\rangle + ...
\end{equation}
The final state with the lowest energy is associated with the main
2p photoelectron line, while the other higher energy states give
rise to the satellite lines, although in all cases a strong mixture
of orbital states may be present. In particular, for the case of a
filled 3d shell ($n=10$, as in Cu$_2$O), charge transfer to the 3d
orbital is precluded, and the final state is dominated by the
$sp$-screened state, $|5p^53d^{10}\rangle$, with no satellite
peaks.\cite{GTE+88} In addition, multiplet splitting of the final
states due to exchange interaction with the core hole may introduce
further features in the
spectra.\cite{RWG71,VSC76,LWHS81,VP85,GTE+88,OKT92,MUN96,Hufner03}
Strong satellite structures occur predominantly in transition-metal
and rare-earth cations, and it is generally accepted that charge
transfer peaks give the most intense contribution, while multiplet
splitting adds to the fine structure, although separation between
these contributions is sometimes
difficult.\cite{RWG71,VSC76,Hufner79,VP85,MUN96}

For our purposes, we are interested in the satellite features that
may allow us to identify the ionic state of the cobalt cations,
particularly at small cobalt oxide thicknesses, where no satellites
are observed. In a simplified relaxation model, the absence of
satellite features can be understood as a consequence of an
electronic structure consisting of filled valence levels to which
charge transfer from the ligand is
precluded.\cite{VP85,GTE+88,KRK+09} This is expected to be the case
for octahedrally coordinated trivalent cobalt, where the crystal
field splits the 3d levels into a low energy triplet $t_{2g}$ level,
and to a high energy $e_g$ doublet, leading to full occupancy of the
$t_{2g}$ states, with a correspondingly low spin state
($S=0$);\cite{Roth64,Goodenough71,WHMS04} one example where this
occurs is in LaCoO$_3$, whose Co 2p XPS spectrum is shown in
Fig.~\ref{fig:XPS}. In cluster theory language, this corresponds to
the situation where the charge transfer energy $\Delta$ (difference
between the excited $|2p^6 3d^{7}\underline{L}\rangle$ and ground
$|2p^6 3d^{6}\rangle$ initial states of the neutral atom), is larger
than the core-hole--d-electron Coulomb energy, $Q$, such that the
screened $|2p^5 3d^{7}\underline{L}\rangle$ final state remains
higher in energy than, and little hybridized with, the unscreened
$|2p^5 3d^{7}\underline{L}\rangle$ state.\cite{LWHS81,ZWS86,GTE+88}
A similar situation occurring in Fe$^{2+}$ compounds has been
analyzed by Kroll {\it et al}.\cite{KRK+09} Hence, in \cobaltite,
the strongly suppressed satellite peak is explained by the fact that
the octahedrally coordinated Co$^{3+}$ states do not contribute to
charge transfer and therefore to shake-up processes; the remaining
1/3 Co cations are tetrahedrally coordinated Co$^{2+}$ and give rise
to shake-up peaks, since the crystal field now leads to a low energy
$e_g$ doublet and a partially filled higher energy $t_{2g}$
triplet.\cite{Roth64,Goodenough71} A comparison of the binding
energies of the 2p Co edge peaks for several cobalt oxide compounds
(Table~\ref{tab:XPS_Co}, where we excluded mixed valency oxides
other than \cobaltite) supports the view that these satellite
features may be employed to identify the valence state of cobalt in
oxides,\cite{BGD75,OH76,BS83,LAS+08} although exceptions are
apparent, including the presence of (strongly suppressed) satellite
peaks in the layered LiCoO$_2$ (where the Co$^{3+}$ occupy
octahedral sites).\cite{Oku78,KC90,MFKL07} For the listed spinels
where Co$^{2+}$ occupy tetrahedral sites, CoAl$_2$O$_4$,
CoCr$_2$O$_4$, and the tetragonal CoMn$_2$O$_4$ (stable only at
elevated temperatures with parasite phases, including MnCo$_2$O$_4$,
known to develop at ambient temperatures),\cite{Buhl69} the
satellite peak splitting is similar to that of divalent cobalt in an
octahedral environment, and here \cobaltite\ seems to be the
outlier.

\begin{table}[htb]
\caption{Binding energy (B.E.) and satellite peak splitting (S.S.)
of the Co 2p peaks for selected cobalt oxide compounds, in eV. The
second column refers to the crystal field environment, octahedral
(o) or tetrahedral (t), while the third column indicates the formal
valence state of the Co cation. [*] This work.} \label{tab:XPS_Co}
\begin{ruledtabular}
\begin{tabular}{lcllllll}
 Oxide & Site & Ion & \multicolumn{2}{c}{2p$_{3/2}$} &  \multicolumn{2}{c}{2p$_{1/2}$} & Ref. \\ 
    &   & & B.E. & S.S.  & BE. & S.S. & \\ \hline
 CoO & o & 2+ & 780.5 & 5.9 & 796.3 & 6.7 & [\onlinecite{CBR76}]\\
 CoFe$_2$O$_4$ & o & 2+ & 780.6 & 5.1 & 796.2 & 6.5 & [\onlinecite{GXFL08}]\\
 CoFe$_2$O$_4$ & o & 2+ & 779.9 & 6.2 & 795.7 & 6.2 & [\onlinecite{OH76}]\\
 LaCoO$_3$ & o & 3+ & 780.1 & --- & 795.3 & ---&  [*] \\
 CuCoMnO$_4$ & o & 3+ & 780.0 & --- & 795.0 & ---&  [\onlinecite{BS83}]\\
 \cobaltite & o & 3+ & 779.6 & --- & 794.5 & --- & [\onlinecite{CBR76}]\\
 MnCoZnO$_4$ & o & 3+ & 780.4 & --- & 795.4 & ---&  [\onlinecite{BS83}]\\
 LiCoO$_2$ & o & 3+ & 779.5 & 10.6 & 794.6 & 9.5 & [\onlinecite{Oku78,MFKL07}]\\
 CoAl$_2$O$_4$ & t & 2+ & 781.0 & 5.0 & 796.7 & 6.3 & [\onlinecite{OH76}]\\
 CoCr$_2$O$_4$ & t & 2+ & 780.4 & 5.3 & 796.3 & 6.2 & [\onlinecite{OH76}]\\
 CoMn$_2$O$_4$ & t & 2+ & 780.2 & 5.6 & 796.3 & 6.5 & [\onlinecite{OH76,BS83}]\\
 \cobaltite & t & 2+ & 780.7 & 8.8 & 796.0 & 8.5 & [\onlinecite{CBR76}]\\
\end{tabular}
\end{ruledtabular}
\end{table}

The case of Co$^{3+}$ in a tetrahedral environment should also give
rise to charge-transfer satellite peaks, since in such a crystal
field both the low energy $e_g$ doublet and the higher energy
$t_{2g}$ triplet have empty states.\cite{Roth64,Goodenough71} These
observations suggest that the interfacial oxide layer present at low
cobalt oxide thicknesses consist of octahedrally coordinated
Co$^{3+}$ cations. We envisage two possibilities that can explain
this result: it either corresponds to a fully oxidized cobalt
compound with a corundum structure, Co$_2$O$_3$, where all cations
are (slightly distorted) octahedrally coordinated; or to the
octahedrally coordinated Co$^{3+}$ layer in the \cobaltite(111)
structure, which also would initiate the spinel growth along the
[111] direction. The LEED and RHEED patterns for the 1.7 and 3.4
\AA\ films, similar to those of the \sapphire\ substrate, seem to
support the first interpretation of the spectroscopy data.

\begin{figure}[t!bh]
\begin{centering}
\includegraphics*[width=5.6cm]{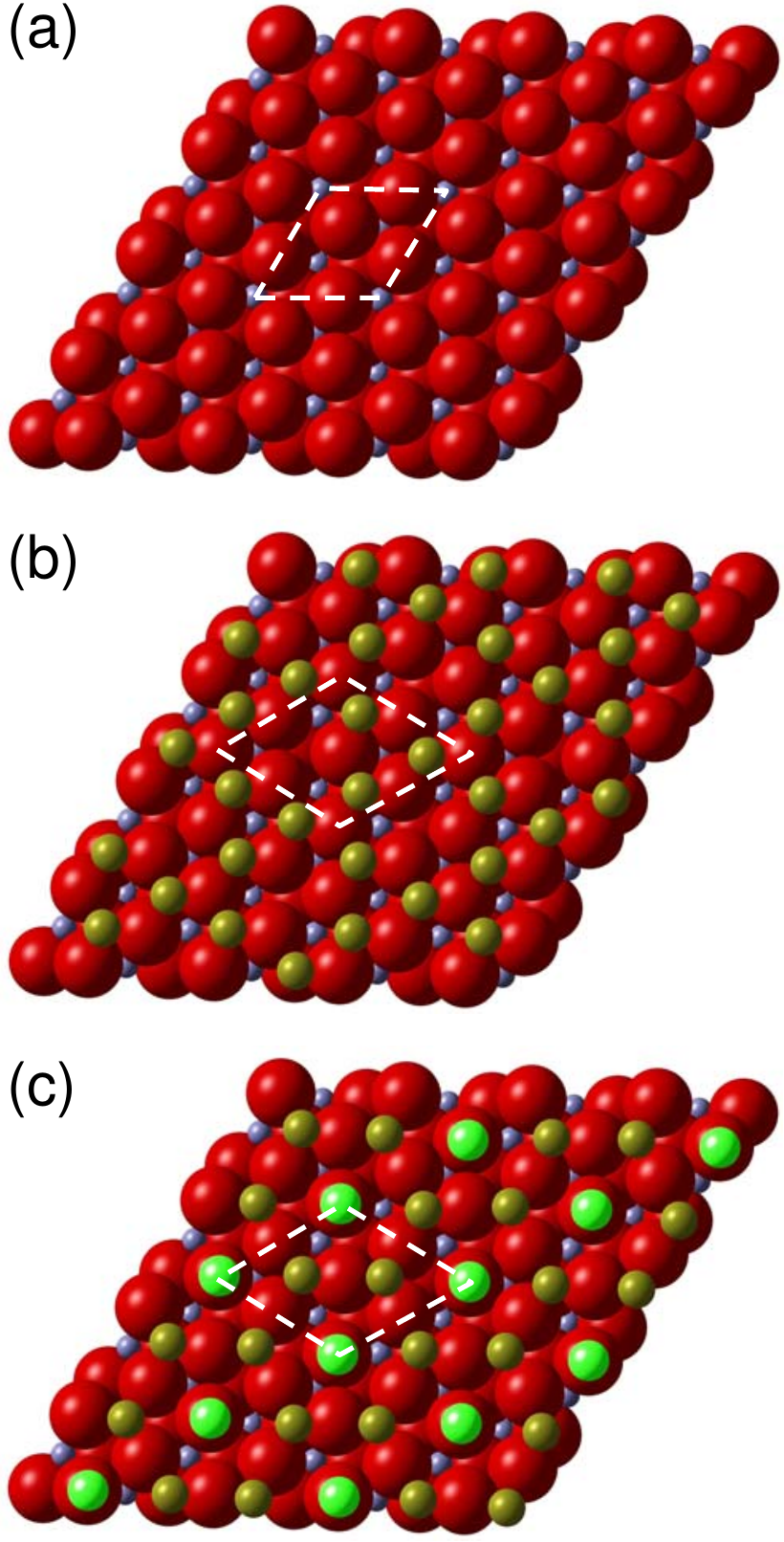}
\caption{(a) Atomic structure of the stoichiometric corundum (0001)
surface. (b) Model for the epitaxial growth of the \cobaltite(111)
structure on the corundum (0001) surface, showing the formation of
the octahedral Co$^{3+}$ layer. (c) Alternative growth model,
showing the formation of the mixed valence tetrahedral
Co$^{2+}$-octahedral Co$^{3+}$-tetrahedral Co$^{2+}$ layer of
\cobaltite(111) on the corundum (0001) surface. For each surface,
the dashed area indicates the primitive surface unit cell.}
\label{fig:model}
\end{centering}
\end{figure}

Is it now well established that under the preparation conditions
used here, the \sapphire(0001) surface is the non-polar, ($1\times
1$) Al-terminated
surface.\cite{MVG93,AR97,GRBG98,Renaud98,JC01,WAM+06} During Co
deposition, the cobalt oxide layer is expected to continue the
\sapphire\ close-packed oxygen sublattice, with either the [AB]A or
[AB]C stacking sequences, where the square brackets enclose the hcp
stacking of the \sapphire\ O sublattice. The first stacking sequence
is a continuation of the hcp stacking and contains octahedral
intersticial sites only, while the second stacking sequence also
contains tetrahedral sites and is expected to mark the onset of the
spinel crystal structure. Based on the LEED and XPS results, and
considering the atomic configurations expected for the available
cationic interstitial sites in the close-packed O sublattice, we
propose next a model to explain the epitaxial relationships observed
in the spinel(111)/\sapphire(0001) system. In this model, the
transition from the corundum to the spinel structure can occur in
two possible ways, schematically shown in Fig.~\ref{fig:model}.
Starting with the stoichiometric corundum (0001) surface,
Fig.~\ref{fig:model}(a), (i) the Co cations occupy non-corundum
octahedral sites concomitant with the initiation of the fcc stacking
of the O sublattice, forming the octahedral Co$^{3+}$ layer of the
\cobaltite(111) structure, Fig.~\ref{fig:model}(b); or (ii) the Co
cations occupy both corundum sites (which are tetrahedrally
coordinated in the fcc stacking of the O sublattice), and octahedral
and tetrahedral non-corundum sites to form the mixed valence
Co$^{2+}$-Co$^{3+}$-Co$^{2+}$ layer of \cobaltite(111), as shown in
Fig.~\ref{fig:model}(c). Hence, one finds that the spinel structure
can very naturally continue the corundum structure without causing
undue violence to the cationic distribution at the interface between
the two crystal structures. The different possible ways in which the
spinel structure can be generated, and the different equivalent, but
not identical, planes that form the \cobaltite\ unit cell along the
[111] direction, imply that the presence of stacking faults and
antiphase boundaries are very likely. They may be responsible, in
part, for the surface disorder observed in the RHEED and LEED
patterns.

We can, therefore, reconstruct the most likely growth process in the
\cobaltite/\sapphire(0001) system. We assume in the following that
one unit layer corresponds to the separation between planes of the
close-packed O sublattice, of about 2.3 \AA, using the lattice
parameter of \cobaltite. In broad terms, we envisage the cobalt
oxide film growth as follows. At the earliest stage of growth, Co
cations fill the empty octahedral corundum sites of the non-polar
\sapphire(0001) surface, resulting in a corundum-like layer composed
of half Al$^{3+}$ cations and half Co$^{3+}$ cations. At 3.4 \AA,
the cobalt oxide film corresponds to about 1.5 atomic layers, and
the observation of octahedrally coordinated Co$^{3+}$ cations in XPS
at this thickness could result from the formation of the
octahedrally coordinated plane of \cobaltite\ upon the onset of the
oxygen C layer of the fcc stacking. However, the LEED patterns show
no evidence for the formation of the spinel (111) surface up to
about 17 \AA. Hence, we conclude that the cobalt oxide film
continues with the corundum structure up to at least 2-3 atomic
layers, forming effectively an interfacial cobalt sesquioxide layer,
mediating the transition between the sapphire and spinel structures.
We do not observe surface reconstructions (LEED, RHEED) or changes
in valency of the Co cations (XPS) during the initial stages of
growth; this may be expected if the corundum structure grows in
Co-O$_3$-Co layer units, which are charge compensated. Since the
Co$_2$O$_3$ phase is not thermodynamically the most stable phase,
the \cobaltite\ phase eventually sets in, with the transition
occurring as described in the model above. This is observed to occur
at 17 \AA, where the XPS shows the presence of tetrahedral Co
cations, characteristic of the spinel \cobaltite(111) structure.

\section{Conclusions}

We have studied in detail the early growth stages of epitaxial
\cobaltite(111) films grown on \sapphire(0001) single crystals. The
RHEED and LEED results show that film growth proceeds via the
Stranski-Krastanov mode; the electron diffraction data, in
combination with spectroscopic characterization by XPS, indicate the
formation of an octahedrally coordinated, fully oxidized,
interfacial cobalt oxide film that mediates the transition between
the corundum and the spinel crystal structures. These results are a
telling example of the strongly modified structural and electronic
properties of metal oxides that are induced by the change in crystal
symmetries at interfaces.

\begin{acknowledgments}
The authors acknowledge financial support by the NSF through MRSEC
DMR 0520495 (CRISP), and by the US DOE Basic Energy Sciences Grants
DEFG02-98ER14882 and DEFG02-06ER15834. The surface crystal
structures were produced with the aid of the DL Visualize public
software.\cite{Searle01}
\end{acknowledgments}


\end{document}